\documentclass[11pt,fleqn]{article}

\usepackage{amsmath}
\usepackage{amsfonts}
\usepackage{graphicx}
\usepackage{dcolumn}
\usepackage{latexsym,slashed}
\usepackage{amssymb,amsfonts}
\usepackage{bm}        %% bold mathematics
\usepackage{amsmath,mathrsfs}   %% non gradito a latex2html!

\newcommand{\address}[1]{\begin{center}\large #1\end{center}}

\def\ba{\begin{eqnarray}}
\def\ea{\end{eqnarray}}

\def\be{\begin{equation}}
\def\ee{\end{equation}}

\catcode`\@=11
\begin{document}
\tolerance=5000

\title{\bf{A note on a  mimetic scalar-tensor cosmological model} }

\author{
Yevgeniya  Rabochaya$\,^{(a)}$\footnote{rabochaya@science.unitn.it},
 and
Sergio~Zerbini$\,^{(a)}$\footnote{zerbini@science.unitn.it}}

\date{}
\maketitle
\address{$^{(a)}$ Dipartimento di Fisica, Universit\`a di Trento,\\
 TIFPA-INFN,  Via Sommarive 14 -- 38123 Povo\\
 Italia. 
}
\medskip \medskip

\begin{abstract}
A specific Horndeski scalar-gravity mimetic model is investigated within a FLWR space-time.
The mimetic scalar field is implemented via a Lagrangian multiplier, and it is shown that  the model has equations of motion formally similar to the original simpler mimetic matter model of Chamseddine-Mukhanov-Vikman.
Several exact solutions describing inflation, bounces, future time singularities  are presented and discussed.

\end{abstract}

%\pacs{ }

\section{Introduction}

It is well known that General Relativity (GR) with additional suitable positive cosmological constant and ordinary radiation and matter, describes quite well the whole history of the Universe, including Dark Energy era.  With an additional scalar degree of freedom, this model may also describe the primordial inflationary period, which is necessary for solving the horizon and flatness problems. This is essentially the so called $\Lambda$-CDM model, or  standard cosmological model, and it has been recently tested with high accuracy \cite{Plank1,Plank2}.  

For a later comparison, we recall that in a flat FLRW space-time
\be
ds^2=-dt^2+a(t)^2 d^2 \vec{x}\,,  
\ee   
and GR plus ordinary matter with an equation of state $p=\omega \rho$,  $\rho$ density of radiation or matter, lead to
 continuity equation 
\be
\dot{\rho}+3H (1+\omega)\rho=0\,,
\label{mc}
\ee
and 
\be
2 \dot{H}+3 H^2 =-\omega \rho\,,
\label{gr}
\ee
where $H(t)$ is the usual Hubble parameter. When the solution of equation (\ref{mc}), 
namely $\rho(t)=\rho_0 a(t)^{-3(1+\omega)}$ is taken into account, one has a non linear second order differential equation for $a(t)$. Of course, in GR, one may make use of the Friedmann equation to arrive directly at the explicit solution for $a(t)$, a textbook result. In presence of a scalar field,  things are not so simple, and, in general, it not easy to find exact solutions (see, for example \cite{sasha1,sasha2}).  

Furthermore, in the standard cosmological model, Dark Matter and Dark energy issues are still under investigation, since some unsolved issues regarding their interpretation are present. In fact, it is well known that Dark Energy effect is well parametrized by the inclusion of a tiny positive cosmological constant, but the coincidence and cosmological constant problems arise. In the physics of elementary particle there exist few candidates for Dark Matter, but experimental verification is stil lacking, and other alternatives are possible.  

For these reasons, in this paper, we would like to consider a generalization of the so called mimetic dark matter-gravity models \cite{muk1,muk2,vik}. This proposal may be considered a minimal modification of GR, in which cosmological dark matter may be described. Soon  it has also  been realized \cite{Derue,barvi} that this class of models are related to GR by singular disformal transformations.  Disformal transformations were introduced by Bekenstein \cite{beke}. He was able to show that  as a consequence of the diffeomorphism invariance of GR, any metric tensor $g_{\mu\nu}$ may be parametrized by a fiducial metric  $l_{\mu\nu}$ and by a scalar field $\phi$. As a consequence, $\phi$ seems to describe a new gravitational degree of freedom, but  as soon as the disformal transformation is invertible,  it turns out that no additional degrees of freedom are present, and GR is recovered.  However, if the disformal transformation is singular, then the scalar $\phi$ becomes a new degree of freedom. This is a general fact \cite{mata}. We will not make use of this very powerful approach and we refer to original papers. In fact, another equivalent approach is possible \cite{golo,barvi,mata}, and within this approach  a  Lagrangian multiplier is introduced \cite{Lim,Capo} in order to implement the ``mimetic''  constraint 
\be
\partial_\mu \phi \partial^\mu \phi=-1\,.
\ee
Very recently,  a  general Horndeski scalar-tensor mimetic  theory has been considered, and  the two possible approaches, the singular disformal transformation and the Lagrangian multiplier approach have been shown to be equivalent \cite{mata}.  Concerning different aspects and generalization of mimetic gravity see, for example \cite{Odi1,Odi2,Odi3,myrza1,myrza2,myrza3}.  

In this paper, in order to deal with mimetic field $\phi$, we shall follow  the Lagrangian multiplier approach.

The outline of the paper is the following. In Section 2, the mimetic model is introduced.  In Section 3 exact solutions in the absence of matter are presented, while in Section 4, matter or radiation is included. The paper ends with the conclusions and an Appendix. We use units in which the reduce Planck mass is $M_P^2=1$. 

\section{Mimetic scalar-tensor  gravity model}

In this Section we start with the following mimetic scalar tensor gravity model 
\be
I =  \int_{\mathcal M} d^4 x \sqrt{-g} \left(\frac{R}{2} + \lambda(X-\frac{1}{2})  -V(\phi) \right)+I_H+I_m\,, 
\label{d}
\ee
where $I_m$ is the usual matter-radiation action and the higher order contribution is given by 
\be
I_H=\int_{\mathcal M} d^4 x \sqrt{-g}\left(\alpha( X R+(\Delta \phi)^2-\nabla_\mu \nabla_\nu \phi \nabla^\mu \nabla^\nu \phi)+\gamma 
\phi G_{\mu \nu} \nabla^\mu \nabla^\nu \phi -\beta \phi  \Delta \phi \right)\,,
\label{h}
\ee
with
$X=-\frac{1}{2}g^{\mu \nu}\partial_\mu \phi \partial_\nu \phi$,  $\lambda$ is a Lagrangian multiplier and $\phi$ is  the mimetic scalar field, $\beta$, $\alpha$, and $\gamma$ are constants. The above Lagrangian is a particular case of the general Horndeski Lagrangian \cite{horn,Deffa,DeFe}. Some examples of Horndeski mimetic gravity models have  been considered recently in \cite{mata}. 

When  the constants $\beta$, $\alpha$, and $\gamma$ are vanishing, and in absence of matter,  the above model reduces to the original mimetic gravity proposed by Chamseddine and Mukhanov \cite{muk1}. When $\alpha$ and $\beta$ are vanishing, the model reduces to the one studied in \cite{NO,Cogno11}, see also \cite{myrza3} for other aspects. 

In order to study the dynamics of above model in a flat FLRW space-time, one has to make use of
\be
ds^2=-e^{2b(t)}dt^2+a(t)^2 d^2 \vec{x}\,,  
\ee
here $b(t)$ is arbitrary dynamical variable which takes the value $b=0$ after variations. The action 
(\ref{d}) can be written as a functional of $a(t)$, $b(t)$ and $\lambda$. Variation with respect to $\lambda$ and  assuming $\phi$ to depend only on $t$ give
\be
\dot{\phi}^2=1\,.
\label{l}
\ee
Thus, in the following, one may take $\phi=t$. Variation with respect to $b$ gives  the generalized Friedmann equation and reads 
\be
3 H^2 (1-3\alpha+3\gamma)-V + \beta-\rho=\lambda  \,.
\label{f1}
\ee
Here $H=\frac{\dot{a}}{a}$, the Hubble parameter, and $\rho$ is the matter-radiation density. 

Making use of the equation of state $p=\omega \rho$,  variation with respect to $a$ leads to
\be
c_1 (2 \dot{H}+3 H^2 )=V+\beta-\omega \rho\,,
\label{f2}
\ee
where
\be
c_1=(1-\alpha+\gamma)\,. 
\ee
Furthermore, the diffeomorphism invariance leads to the continuity equation 
\be
\dot{\rho}+3H (1+\omega)\rho=0\,.
\ee
Finally, the equation of motion associated with $\phi$ is also present, but it is a consequence of the other equations of motion, thus it is trivially satisfied.

A remark is in order. The equation of motion  (\ref{f2}) in this Horndeski mimetic model does not contain the Lagrangian multipliers, and it is similar to the one valid in GR plus ordinary matter. However, here the mimetic potential $V$ appears in a very peculiar way, and this helps a lot in the search for exact solutions. In fact, one is dealing with  a  non linear first order Riccati differential equation. 

Furthermore, as in GR, one may have de Sitter solution $H=H_0$ if and only if the potential is a constant, and $\omega=-1$ or $\rho=0$. Furthermore, one may have de Sitter solution with vanishing potential and absence of matter, but with the constant $\beta \neq 0$. The effective cosmological constant depends on the ratio $\frac{\beta}{c_1}$. Thus,  in presence of a non trivial potential, one may have only quasi-de Sitter solution, and inflation and current acceleration may be described.
With regard to other solutions, there exist several possibilities.
\section{Absence of matter}
In absence of matter, the Riccati equation may be recast in an homogeneous  linear second order differential equation. In fact, introducing the new variable $y=a^{3/2}$, one has $H=\frac{2 \dot{y}}{3 y}$, and 
\be
\ddot{y}-\frac{3 }{ 4 c_1}(V(t)+\beta) y=0\,.  
\ee
The general solution of this kind of equation is not known. Approximate solutions may be investigated by WKB method. Alternatively, one may use another approach, the so called adiabatic invariants method, and for the  sake of completeness we report it in the Appendix. The reconstruction method is also possible and it has been investigated in \cite{Odi2}. In the following, we shall discuss some exact solutions.

As a first  example, let us   consider a quadratic potential
\be
V=-\beta+2V_0+\frac{3V_0^2}{c_1}(\phi-\phi_0)^2\,.
\ee
This choice gives
\be
V(t)=-\beta+2V_0+\frac{3V_0^2}{c_1}(t-t_0)^2\,.
\ee
Then, it is  easy to show that the solution is
\be
y(t)=y_0 ^{\frac{3 V_0}{4 c_1}(t-t_0)^2  }\,,
\ee
and, in term of Hubble parameter, one has
\be
H(t)=\frac{V_0}{c_1}(t-t_0)\,. 
\ee
Since $H(t_0)=0$ and $\dot{H}(t_0)>0$, this is an example of a regular ``bounce'' solution. Solution of this kind for an extension of the Starobinski model has ben found in \cite{Seba}, and for other bounce solutions, see \cite{Myrza4}.

Another bounce solution may be obtained by the following choice of $\beta=-2c_1b^2$ and for the potential
\be
V(\phi)=b^2 c_1 \frac{\sinh^2 b \phi}{\cosh^2 b \phi }\,,
\ee
with $b$ a real parameter. In this case, the bounce solution is
\be
a(t)=a_0 \cosh bt\,,
\ee
with
\be
H(t)=b\frac{\sinh b t}{\cosh bt }\,.
\ee
The bounce is at $t=0$. This kind of bounce solution may be obtained in a specific model of non-local gravity in FLRW space-time \cite{Biswas1,Biswas2}. Here the bounce has been related to a simple potential in the mimetic scalar field. 

Other exact solutions have been presented in 
\cite{muk2}.

\section{Presence of matter}

If matter is present,  one may introduce the e-fold time $N=\ln a$. As a result, the continuity equation becomes
\be
\frac{d \rho}{dN}=-3 (1+\omega)\rho \,.
\ee
with solution
\be
\rho(N)=\rho_0 e^{-(1+\omega)N}\,.
\ee
Furthermore, the equation of motion for $H$ becomes
\be
c_1\left(\frac{d H^2}{d N}+3 H^2 \right)=V+\beta -\omega\rho(N)\,.
\label{r}
\ee
If
\be
c_1=(1+\gamma-\alpha)\neq 0\,,
\ee
the solution is
\be
H^2(N)=e^{-3N}\left( C +\int dN  e^{3N}  \frac{V(N)+\beta-\omega\rho_0e^{-(1+\omega) N}}{c_1} \right)\,,
\ee
where $C$ is an integration constant. Furthermore, if $\omega+1 $ is non vanishing, one has
\be
H^2(N)=Ce^{-3N}+\frac{ \beta}{3 c_1}-\frac{\rho_0 \omega}{\omega+1}e^{-3\omega N}+
\int dN  e^{3N}  \frac{V(N)}{c_1}\,.
\ee
In the case $\omega=-1$, one instead has
\be
H^2(N)=Ce^{-3N}+\frac{ \beta}{3 c_1}+\frac{\rho_0 }{c_1}+
e^{-3N}\int dN  e^{3N}  \frac{V(N)}{c_1}\,.
\ee
Some comments are in order. Since
\be
t(N)=\int \frac{dN}{H(N)}
\ee
we may obtain $N=N(t)$ and $a(t)=e^{N(t)}$. The contribution depending on $C$, is the contribution associated with the mimetic dark matter. The term depending on $\frac{\beta}{c_1}$ acts again as an effective cosmological constant. Another constant contribution may be obtained by the  simplest choice for the potential  $V=V_0$, namely a constant  potential. 
In this  case, we have 
\be
H^2(N)=Ce^{-3N}+\frac{ \beta}{3 c_1}-\frac{\rho_0(1+\omega)}{\omega}e^{-3(1+\omega)N}+ \frac{ V_0}{ c_1}\,.
\ee
and
\be
H^2(N)=Ce^{-3N}+\frac{ \beta}{3 c_1}+\frac{\rho_0 N}{c_1}e^{-3 N}+ 
\frac{ V_0}{ c_1}\,.
\ee
For consistency, we have to assume $\beta+V_0 >$. Thus, these  solutions tend for large $N$ to the de Sitter space-time.

Another interesting example of potential is the one satisfying
\be
V(N)=3g(N)+\frac{d g}{dN} 
\ee
with $g(N)$ a known function. In this case, the solution is
\be
H^2(N)=Ce^{-3N}+\frac{ \beta}{3 c_1}-\frac{\rho_0(1+\omega)}{\omega}e^{-3(1+\omega)N}+ 
\frac{ V_0}{ c_1 }g(N) \,.
\ee
As an example, take
\be
V(N)=V_0 (N-N_0)^{b -1}\left(3(N-N_0)+b  \right) \,,
\ee
with $b \neq 2$. As a result
\be
H^2(N)=Ce^{-3N}+\frac{ \beta}{3 c_1}-\frac{\rho_0(1+\omega)}{\omega}e^{-3(1+\omega)N}+ 
\frac{ V_0}{ c_1 }(N-N_0)^{b} \,.
\ee
The nature of the solution may depend on the sign of $b$. To simplify the discussion, assume $C=\beta=\rho_0=0$, and $\frac{ V_0}{ c_1} >0$  Thus
\be
H(N)=\left(\frac{ V_0}{ c_1 }\right)^{1/2}(N-N_0)^{b/2} \,.
\ee
As a consequence
\be
t-t_0=\left(\frac{ V_0}{ c_1 }\right)^{-1/2} \frac{2}{2-b}(N-N_0)^{1-b/2}\,,
\ee
and
\be
N-N_0=\left(\frac{ V_0}{ c_1 }\right)^{1/(2-b)}\left(\frac{2-b}{2} (t-t_0)\right)^{2/(2-b)}\,.
\ee
The Hubble parameter reads
\be
H(t)=A (t-t_0)^{b/(2-b)}\,.
\ee
If $b >2$, as well as for $b <0$, there exists a future-time singularity  (see, for example \cite{Noji,Acqua} and references therein). If $0 <b< 2$, there is a bounce solution. For example, for $b=1$, one has the bounce solution
\be
H(t)=\frac{V_0}{c_1} (t-t_0)\,.
\ee
In this case the potential is 
\be
V(\phi)=V_0 \left(1+\frac{3 V_0}{ c_1 }(\phi- \phi_0)^2   \right)\,,
\ee
in agreement with the result discussed in  Section 3.

\section{Conclusions}
In this paper, a specific cosmological Horndeski scalar-gravity mimetic model has been  investigated within a FLWR space-time. The mimetic scalar field has been  implemented making use of a Lagrangian multiplier, and it has been shown that the model leads to  equations of motion
formally similar to the original simpler mimetic matter model \cite{muk1}.
Several exact solutions  describing inflation, bounces, future-time singularities  have been presented and discussed. 

It should  be interesting to investigate  spherically symmetric static solutions of this generalized mimetic model along the lines of reference \cite{Myrza5}.   

\section{Appendix}

In the following we briefly discuss an alternative method to find solution of a differential equation of the kind 
\be
\ddot{y}+Q(t) y=0\,.  
\label{a}
\ee
The above linear homogeneous differential equation can be associated with  the following non linear differential equation, dubbed Ermakov-Pinney equation  (see \cite{Lewis1,Lewis2,gray}), namely
\be
\ddot{u}+Q(t) u=\frac{h^2}{u^3}\,.  
\label{b}
\ee
In fact, the following  result  holds true; the solution $y$ and $u$ are related by
\be
y=u \sin \theta\,, \quad \theta=\int \frac{h}{u^2}dt \,, 
\ee
and the constant $h$ is given by the so called Lewis adiabatic invariant
\be
h^2=\frac{h^2 y^2}{u^2}+(u\dot{y}-\dot{u}x)^2\,.
\ee
As an example, let us consider
\be
Q(t)=\frac{q}{t^4}\,, \quad q>0\,.
\ee
Then, it is easy to show that an exact solution of the Ermakov-Pinney equation is
\be
u(t)=(\frac{h^2}{q})^{1/4} t\,,
\ee
and angle $\theta(t)$ reads
\be
\theta(t)=\theta_0-\frac{q^{1/2}}{t}\,.
\ee
As a result, the exact solution is 
\be
y(t)=(\frac{h^2}{q})^{1/4} t \sin (\theta_0-\frac{q^{1/2}}{t} )\,,
\ee
a non trivial result. Thus, we also  obtain an exact solution associated with a singular potential 
\be
V(\phi)\equiv \phi^{-4}\,.
\ee 
Other exact solutions of the Ermakov-Pinney equation can be found in \cite{Lewis1,Lewis2,gray}.

\section{Acknowledgments}
 We wish to thank G.\ Cognola, L.\ Sebastiani, and S.\ Vagnozzi for valuable discussions.

\end{document}